\documentclass[aps,prl,twocolumn,superscriptaddress,nofootinbib,nobalancelastpage,showpacs,nobibnotes]{revtex4}
\usepackage{epsfig}

\begin{document}
\title
{``Classical'' instabilities and ``quantum" speed-up in the evolution of neutrino clouds.}

\author{R. F. Sawyer}\email{sawyer@vulcan.physics.ucsb.edu}
\affiliation{Department of Physics, University of California at
Santa Barbara, Santa Barbara, California 93106}

\begin{abstract}
We study some examples of collective behavior in neutrino clouds governed by the
neutral-current neutrino-neutrino interaction.  The standard equations for analyzing such systems
are rederived in a two-step process: first, a replacement of the full interaction Hamiltonian with
a ``forward" Hamiltonian that contains only the momentum states that were initially occupied by a
neutrino of one flavor or another; second, a factorization assumption that reduces the time evolution
problem to the solution of coupled equations for the expectations of various bilinear forms in the 
neutrino fields. We designate the latter as the ``classical" equations. We analyze some 
solutions of these equations in cases in which the initial momentum and flavor distributions of
neutrinos are strongly anisotropic in space. In some cases we find an instability that leads
to rapid evolution of the flavor-angle distribution, even when it is seeded by a very small
initial flavor mixing (or alternatively by a very small neutrino mass$^2$ term).
Turning to the more complete case in which we do not assume the classical factorization, but instead 
solve for the evolution under the influence of the full ``forward" Hamiltonian, we find
the possibility of rapid evolution, under our definition, even when there is no seeding from conventional 
neutrino mixing. This ``speed-up", which occurs in exactly the same parameter range as do the
instabilities in the classical case, to some degree confirms earlier conjectures of such collective
speed-up's. However the time scales found in the present work are larger than those 
previously conjectured time scales by a factor
of order $\log N$, where $N$ is the number of neutrinos in a volume of dimension of the
reaction length. Even with this lengthening of the ``speeded-up" scale there can be situations in which
these effects will dominate the short term behavior of a system.

\pacs{95.30.Cq, 97.60.Bw}

\end{abstract}

\maketitle

\section{1. Introduction}

Clouds of neutrinos, within media that may or may not have other constituents,
 are capable of behaving in many different interesting ways,
some of which may even be relevent to real astrophysical situations. Within the context of the
theory in which the neutrinos have standard-model couplings to baryons and leptons, augmented by
a flavor-mixing mass term, we enumerate some possible behaviors that go beyond
vacuum neutrino oscillations:
\begin{enumerate}
\item  
 The familiar index of refraction or ``matter" effect from forward scattering on electrons \cite{wolf}, \cite{ms} 
which enters the interpretation of the results of some solar neutrino experiments.
\item
Slowing of oscillation rates in the presence of ``flavor-measuring" noise \cite{slow1}-\cite{slow3}.
\item Synchronization of oscillations in the presence of ``flavor-blind" noise,  \cite{BSV}.
\item Synchronization of oscillations through the nonlinear effects of the
neutral-current neutrino-neutrino interaction \cite{s1}-\cite{s8}.
\item Many-body effects that may \cite{bell}, \cite{rfs}, or may not \cite{fl}, speed-up
some neutrino reaction rates.
\end{enumerate}
In the present paper we discuss yet another set of behaviors, closely linked
to items \#4 and \#5, above, but with potential physical effects that are distinct.
To set the context we note  three different time-scales that can enter a 
system in which only neutrinos are present.

a) $T_{\rm scat}=(G_F^2 E^2 n_\nu)^{-1}$, the characteristic time for
ordinary scattering of
a neutrino in a cloud, where $E$ is the energy scale of the neutrinos in the cloud
and $n_\nu$ is the neutrino number density. In what follows we consider only times
that are much less than $T_{\rm scat}$.

b) $T_S= (G_F n_\nu)^{-1} $, the time over which the phase of the wave function
of a neutrino is changed by an amount of order $2 \pi$, due to interactions with
other neutrinos in the cloud. We refer to $T_S$ as the ``short time scale".

c) $T_L = (\delta m^2/E)^{-1}$, the neutrino oscillation time scale, which we designate
as the ``long time scale".
\newline

Under all conditions that we shall consider, which include the parameter
domains for both early-universe and supernova applications, we will have
$T_{\rm scat}>>T_L>>T_S$. In this case, the evolution over short time scales will be determined
by a ``forward" neutral current interaction Hamiltonian, which we define in detail later.
The evolution over the neutrino
oscillation time-scale is determined by a combination 
of the neutrino-mass terms and the neutral-current terms.
These two effects are combined in a nonlinear equation for the neutrino
distribution function that has been presented in refs. \cite{sr}, \cite{prs} and elsewhere and has been discussed and
applied by numerous authors \cite{sn}-\cite{raff}. In certain circumstances the result is the synchronization
of neutrino oscillation modes, as mentioned in \# 4 in the above list.
In the applications of these nonlinear evolution equations, whether or not synchronization is operative,
authors have usually assumed
an isotropic medium. Because the forward scattering of neutrinos is strongly
dependent on the angle of incidence, the equations for the anisotropic case are more complex.

Here we study some examples of violently non-isotropic cases, and find behavior 
that is qualitatively very different from that in isotropic cases. For example, in the isotropic case, if
the system begins in a near flavor-eigenstate, with a small flavor mixing as a perturbation,
then there will ensue slow oscillatory behavior, with a time scale $T_L$. In some non-isotropic examples, on the other hand, we will find exponential
growth of the perturbation, on the short time scale, $T_S$, followed by very non-linear and large
amplitude oscillations, still on the short time scale. These oscillations do not change
the total numbers of neutrinos of a particular flavor in a system, nor the all-over energy
distribution or angular distribution, but they can quickly
change the angular and spectral distributions of a particular flavor. When we explore, to some degree,
the space of combined anisotropy and initial arrangement of flavors, we find that there
are stable and unstable regions and that the boundaries will be rather hard to classify.
These effects are surely of interest with respect to supernova physics, and could conceivably
be of interest with respect to deviant theories of early universe evolution.

Following many previous studies, we begin with a system that contains two flavors of
neutrinos and with equations that capture the complete effects
of forward scattering. We make our considerations absolutely
concrete by beginning with a ``totally forward" effective Hamiltonian, $H_{\rm for}$,
which contains both neutrino oscillations and the neutral current
interactions.  We expand the neutral-current Hamiltonian in terms of the creation and annihilation
operators $(a^{r}_i )^\dagger ,a^{r}_i ,(a^{s}_i) ^\dagger ,a^{s}_i$ for neutrinos of flavors
$r$ and $s$. Taking the set of momentum states $\{p_i \}$ that are 
occupied in the initial state, by neutrinos of either species $\nu_r$, $\nu_s$ (or mixtures thereof) and dropping
from $H$ all of the operators referring to other momenta, we obtain
\begin{equation}
H_{\rm for}=\sum _i \lambda_i \vec B \cdot \vec \sigma_i+{G_F \over 2 \sqrt{2} V} \sum_{i,j}d_{i,j}
\vec \sigma_i \cdot \vec \sigma_j+H_R~,
\label{ham}
\end{equation}
where the operators $\vec \sigma_i$ are given by,
\begin{eqnarray}
  \sigma^{(1)}_j=(a^r_j)^\dagger a^s_j + (a^s_j)^\dagger a^r_j \,\,,
\nonumber\\
\sigma^{(2)}_j =-i(a^{r}_j)^\dagger a^{s}_j +i (a^{s}_j)^\dagger a^{r}_j\,\,,
\nonumber\\
\sigma^{(3)}_j=
(a^{r}_j)^\dagger a^{r}_j - (a^{s}_j)^\dagger a^{s}_j \,\,.
\label{operators}
\end{eqnarray}
Using the usual anticommutation rules for the operators $a$, $a^\dagger$ yields the commutation rules of Pauli spin 
operators for the $\vec \sigma_i$. The coefficients $d_{i,j}$ are given by,
\begin{equation}
d_{i,j}=1-\vec p_i \cdot \vec p_j /(|\vec p_i|\, |\vec p_j|).
\end{equation}
The neutrino oscillation parameters are contained in $\lambda_i=\delta m^2 /p_i$, providing the scale, and in the vector $\vec B$, which we take to have unit length.

The residual part of  (\ref{ham}), $H_R$, contains only terms that commute with all of the operators
$\vec \sigma_i^{(\alpha)}$; these consist of the kinetic energy terms and all terms involving only the operators
$(a^r_j)^\dagger a^r_j + (a^s_j)^\dagger a^s_j $; the contribution of these terms to the amplitudes that we
calculate will be all-over phase factors. We will occasionally use the language
of spins on sites to characterize the states of the system governed by (\ref{ham}), while understanding that the
sites are really momentum states and the spins are really flavors.
We write the
Heisenberg equations of motion for the operators $\vec \sigma_i(t)$,
\begin{equation}
{d \over dt} \vec \sigma_i=2 \lambda_i \vec B \times \vec \sigma_i - {\sqrt{2} G_F \over   V}\sum_j d_{i,j}\,\, \vec \sigma_i \times \vec \sigma_j \, .
\label{eom}
\end{equation}

All of the results reported in refs. \cite{s7}-\cite{raff} depend on the following two approximations :

(a.) Replacement of the equation of motion (\ref{eom}) by its expectation value, $\langle \rangle$, 
with the assumption that, on the RHS, we can make the replacement,

\begin{equation}
\langle \vec \sigma_i \times \vec \sigma_j \rangle \rightarrow 
\langle \vec \sigma_i \rangle \times \langle \vec \sigma_j \rangle\, ,
\label{uncor}
\end{equation}
as though the spins on different sites are not correlated. We shall refer to the resulting equations as
``classical"; having used the Heisenberg commutator to derive the equations, we then treat the
expectation values as classical variables.
In what follows, we take the initial wave function of the system to embody the
assumption (\ref{uncor}). But  (\ref{uncor}) does not continue to hold as time progresses. 
Arguments have been made in the literature, albeit implicitly, that
in the limit in which the number of particles $N \rightarrow\infty$ equations like (\ref{uncor})
will be applicable over a large time scale, classical behavior being expected
for large ``spins". We shall examine the validity of this assumption later.

(b.) Isotropy of all distributions. This appears to allow us to make the replacement $d_{i,j}\rightarrow 1$.
At the very least, the applicability of this assumption depends on the application that is being pursued.
Clearly, and as noted, e. g.,  in ref. \cite{sig}, supernova applications require the retention of the angular
dependence embodied in the complete $d_{i,j}$'s. In (nearly) isotropic early-universe applications
it is probably adequate to use $d_{i,j}\rightarrow 1$, as in refs. \cite{earlyu}; however, it is conceivable
that instabilities of non-istropic distributions in the (nearly) collisionless neutrino sector could combine 
with the fluid dynamics of the collisional component of the matter to produce interesting phenomena.

In the main body of this paper we 
shall adopt assumption (a.), represented by (\ref{uncor}), but drop the assumption (b.) of isotropy. We return in sec 6 
to examination of the
limitations of assumption (a.).

In  (\ref{eom}), the operators,  $\sigma_i$, act on the wave functions of individual particles and there
is an extensive factor of volume. To introduce operators with equations of motion and initial
conditions that involve intensive variables, or densities in the present
case, we divide the solid angle for momentum directions into a set of regions $\{a,b...\}$ each with a reasonably
definite direction. We can also define energy bins that further distinguish these regions, in order to
study energy-spectrum-flavor connections. We define the operators
\begin{equation}
 \vec P_a={1\over n_\nu V}\sum_{i \subset \{ a\} }\langle \vec \sigma_i \rangle \, .
\end{equation}
Then we can write the equations of motion (\ref{eom})
as
\begin{equation} 
{d\over dt} {\vec P_a}=\lambda_a {\vec B}\times {\vec P_a}- g   \vec P_a
\times  \sum_{a,b} (1-\cos \theta _{a,b}) \vec P_b
\label{eom2}
\end{equation}
where  $g=\sqrt{2}n_\nu G_F$ and $x_{a,b}=\cos \theta _{a,b}$ is the cosine of the angle between the momentum directions in the two groups. The angle-flavor-energy joint distributions of neutrino density are 
now determined by the $\vec P_a$. 
For example, the respective neutrino densities in the two flavor states $r$ and $s$ are given by
\begin{eqnarray}
n_r=n_\nu (1+ \sum _a P_a^{(3)})/2,
\nonumber\\
n_s=n_\nu (1- \sum _a P_a^{(3)})/2 \, .
\end{eqnarray}

We note that in the isotropic case of assumption (b), we can make the replacement
$\sum_b (1-x_{a,b} )\vec P_b\rightarrow \vec J$ where $\vec J=\sum_a \vec P_a$. In this form the equations (\ref{eom2})
have been studied extensively, both as a curiosity and as an essential tool in studying the evolution
of the early universe in the case of large initial neutrino-anti-neutrino asymmetries.
The qualitative outcome, in this isotropic case, can be stated
fairly simply: If we start with the individual neutrinos in flavor eigenstates, then the timescale 
for evolution of the macroscopic properties is of order $T_L$, just as it would be in the absence of the
neutral current interactions. There is a much discussed effect of the neutral current interactions during
this time period, namely the synchronization of oscillation frequencies. But this synchronization
does not change the general evolution time-scale for any distribution of neutrinos that is reasonably
clustered in energy and that begins from a flavor eigenstate.

In the present paper we consider some solutions of (\ref{eom2}) that begin 
with non-isotropic distributions. We find cases that give fast evolution of macroscopic
properties, in constrast to the behavior described above. The non-linear equations (\ref{operators}) 
for the non-isotropic case with unequal $d_{i,j}$ 
are so complex that we have little understanding of the full range of possibilities. 
However, below we will develop a series of special cases,
which perhaps can serve as a basis for some educated guesses about the broader issues.
In each of these cases we seek interesting behavior involving macroscopic changes, but on the
\underline{fast} time-scale $T_S$. We shall also drop the first term on the RHS of (\ref{eom2})
in favor of an initial condition corresponding to the first (flavor-mixing) term having been turned on (by itself)
for a brief period time, leading to a tiny rotation in flavor space, so that we no longer have
quite an eigenstate of the original flavor operators. We explain later why this is an illuminating 
way to proceed.

\section{2. Two groups of neutrinos with standard couplings}

By a group, we mean an assemblage of neutrinos, each initially in the identical flavor state
and all moving in the same direction, in a ``laboratory" frame in which we do the calculation.
A distribution of energies within a group is allowed. When we have two groups, $A$ and $B$, we can assemble
the operators for the first into a collective operator $\vec \tau=\sum_A \vec \sigma_i$ where the
notation indicates a sum only over group $A$, and a collective operator $\vec \zeta=\sum_B \vec \sigma_i$.
The Hamiltonian (\ref{ham}), discarding $H_R$, and taking the contribution from the neutral current part only
is now,
\begin{equation}
H_{\rm for}={G_F   (1-x_{A,B}) \over \sqrt 2 V }\vec \tau \cdot \vec \zeta \, .
\end{equation}
 Note that the angular
factors, e.g. $1-x_{A,A}=0$, eliminated all interactions among the neutrinos within a group. As explained above, we will first
use the equations for the operator expectations $\vec P_A=\langle \vec \tau \rangle /(n_\nu V)$, 
 $\vec P_B=\langle \vec \zeta \rangle /( n_\nu V)$ 
as in (\ref{eom2}), viz,
\begin{equation}
{d \over dt} \vec P_A =- {d \over dt}\vec P_B =- g(1-x_{A,B})\vec P_A \times \vec P_B \, .
\label{eom3}
\end{equation}
We take initial conditions such that $\vec P_{A,B}^{(3)}=  c_{A,B}$ and
$P_{A}^{(1),(2)}=0$, $P_B^{(1)}=\epsilon_1$ $P_B^{(2)}=\epsilon_2$,
where the coefficients $c_{A,B}$ are of either sign and
of order unity, and where $\epsilon_{1,2}$ are very small. Then we ask whether there are significant changes
of occupancies $P_{A,B}^{(3)}$ in  the shorter time scale $T_S$. The system (\ref{eom3}) conserves 
$\vec P_A(t) +\vec P_B(t) = \vec \alpha$, so that (\ref{eom3}) can be written as, 
\begin{equation}
{d \over dt} \vec P_A(t)= -g (1- x _{A,B}) \vec P_A(t) \times \vec \alpha \, ,
\label{precess}
\end{equation}
with the initial condition $\vec P^{(i)}_A(0)=c_A \delta_{i,3}$. We see that $\vec P_A$ precesses around the
nearly parallel vector $\vec \alpha$ with a rate that is of order $\epsilon T_S^{-1}$. Thus by
our definition of short-term, the short-term change in $P$ is small because of the factor $\epsilon$. 
In the case in which $|c_A-c_B |>> \epsilon$ the change in $P$ is doubly inconsequential, since then $\vec P_A$ is almost parallel to $\vec \alpha$ and the amplitude that oscillates is also small.

\section{3. Two groups of neutrinos with slightly altered couplings}

If we slightly alter the Hamiltonian for the
above case in a way that destroys the SU2 symmetry in the internal space we find an interesting generic phenomenon, one 
which will serve as the prototype for similar phenomena in cases with three or more groups that interact through the
\underline{correct} neutral current coupling.
We take
 \begin{equation}
H_{\rm for}={G_F   (1-x _{A,B} )\over \sqrt 2 V }( \tau^{(1)}  \zeta ^{(1)}+\tau^{(2)}  \zeta ^{(2)}+
\gamma \tau^{(3)}  \zeta ^{(3)}) \, ,
\label{ham4}
\end{equation}
where the SU2 symmetry is broken when $\gamma \ne 1$.
We write the equations of motion for the $P$'s in terms of a rescaled time coordinate, $t'=g[1-x_{A,B}]t$,
\begin{eqnarray}
{d \over dt'} P_A^{(1)}=P_A^{(3)} P_B^{(2)}-\gamma P_A^{(2)} P_B^{(3)},
\nonumber\\
{d \over dt'} P_A^{(2)}=-P_A^{(3)} P_B^{(1)}+\gamma P_A^{(1)} P_B^{(3)},
\nonumber\\
{d \over dt'} P_A^{(3)}=P_A^{(2)} P_B^{(1)}- P_A^{(1)} P_B^{(2)},
\label{eom4}
\end{eqnarray}
with three more equations generated by $A \leftrightarrow B$. 
Introducing the new variables,
\begin{eqnarray}
&x=2P_A^{(3)}\,\,\, ; \,\,\,y=(P_A^{(1)})^2 +(P_A^{(2)})^2+ (P_B^{(1)})^2 +(P_B^{(2)})^2 ,
\nonumber\\
&z=2(P_A^{(2)} P_B^{(1)}-P_A^{(1)} P_B^{(2)}), 
\nonumber\\
&w=2(P_A^{(1)} P_B^{(1)}+P_A^{(2)} P_B^{(2)}),
\end{eqnarray}
we consider only solutions in which the conserved quantity,  $P_A^{(3)}+P_B^{(3)}=0$. For this case (\ref{eom4})
yields the following closed set of equations, where $\dot x\equiv (d/dt')x$, etc.,
\begin{eqnarray}
\dot x=2z\,\,\,; \,\,\, \dot y =-2 x z ,
\nonumber\\
\dot z =-2 x(y+\gamma  w)\,\,\,;\,\,\,\dot w =2 \gamma x z \, .
\label{xyzeqs}
\end{eqnarray}
Once two constants of motion are set, 
\begin{eqnarray}
y+w/\gamma=c_1\,\, ; \,\, y=-{x^2\over 2}+c_2 \, ,
\end{eqnarray}
we obtain the second order equation,
\begin{equation}
{ \ddot x \over 2 } =x^3 (1-\gamma^2)-x[2c_2(1-\gamma ^2)+2 \gamma^2 c_1] ~.
\label{singleeq}
\end{equation}
We take an initial configuration in which $x=1$, $y=\epsilon$ with $\epsilon<<1$, $z=0$, $w=0$, giving
$c_1=\epsilon  $, $c_2=1/2+\epsilon$, $\dot x (0)=0$. This corresponds, for example, to 
beginning with group $B$ entirely in the $s$ flavor state, and with group $A$ in the state derived from
the $r$ flavor state by rotation through an angle $\approx  \sqrt \epsilon $. 
We show in fig. 1 the results of the solution of (\ref{singleeq})  for the case of $\epsilon=.0001$
for various values of the parameter $\gamma$. 
\begin {figure}[ht]
    \begin{center}
       \epsfxsize 2.75in
        \begin{tabular}{rc}
           \vbox{\hbox{
$\displaystyle{ \, { } }$
               \hskip -0.1in \null} 
} &
            \epsfbox{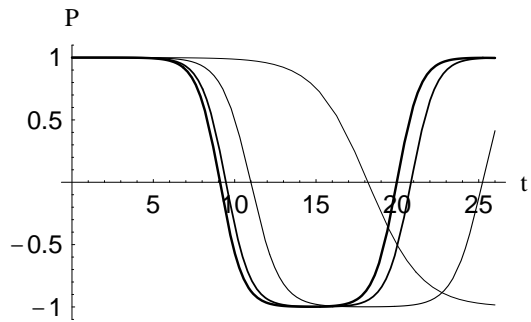} \\
            &
            \hbox{} \\
        \end{tabular}
    \end{center}
\label{fig.1}
\protect\caption
    {%
The flavor turnover of group A, from the solution of eqs.(\ref{eom4}) plotted against the scaled time ($t'$ in text).
$P$ represents 2$P^{(3)}_A$ in text, with $P=1$ indicating 100\% $r$ flavor for group A and
$P=-1$ representing 100\%  $s$ flavor. The parameter $\epsilon$ is set at $.0001$, with curves plotted
for values $\gamma=0,.3,.6,.9$, the heavier lines standing for the lower values of $\gamma$. When $ \gamma \ge 1.0$
the curve is indistinguishable from the line $P=1$.
 }
\end {figure}
We see the complete and abrupt trade of flavors 
between the $A$ group of states and the $B$ group of states at a time that is nearly independent of $\gamma$ for the 
range $0<\gamma <.7$. As $\gamma$ approaches unity
the turnover point starts to recede. For $\gamma \ge 1$ the turnover phenomena 
disappears, the flavors remaining stuck essentially on their original values over periods of scaled time 
$t' < \epsilon^{-1}$, as indicated by the solution (\ref{precess}) for this case. 

Qualitatively one can understand this turnover behavior, for example in the case $\gamma=0$, by observing
that when $\epsilon=0$,  (\ref{singleeq}) becomes just
$\ddot x/2=x^3-x$, which has the familiar kink solution $x={\rm tanh}(t-t_0)$ (as well as the solution $x=\pm 1$).
Our above initial values for the case of very small $\epsilon$ then tie to a periodic array of well-
separated up-kinks and down-kinks, centered at the points $t_n$. Locally, each kink is \underline{nearly} 
of the form ${\rm tanh} (t-t_n)$. However, we are hear interested only in determining the time 
elapsed up until the first turnover. In fig. 2 we show the time development over roughly this shorter span, for a series of values of $\epsilon$'s equally spaced in 
$(-\log \epsilon)$ as the values of $\epsilon$ are successively reduced.
 
\begin {figure}[ht]
\begin{center}
       \epsfxsize 2.75in
        \begin{tabular}{rc}
            \vbox{\hbox{
$\displaystyle{ \, { } }$
              \hskip -0.1in \null} 
} &
           \epsfbox{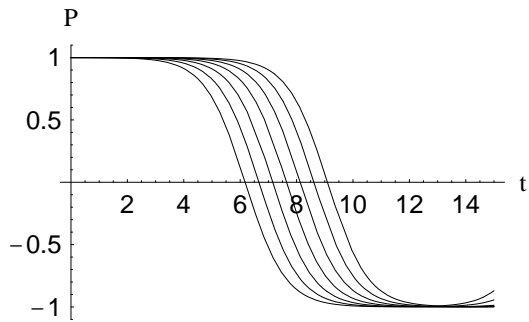} \\
           &
            \hbox{} \\
        \end{tabular}
    \end{center}
 \label{fig.2}
\protect\caption
    {%
As in fig.1, now showing the set of curves for $\gamma=0$, with values $\epsilon=.00001$ times $1,2,4,8,16,32,64$.
The curves farther to the right correspond to the smaller values of $\epsilon$. The equal spacing
is indicative of the $(- \log \epsilon)$ behavior that is one of the principal results of this paper.
 }
\end {figure}
From the equal spacing of the plots we clearly see that the transition time
to the first turnover increases as $(-\log \epsilon)$. We can understand this from the $f(t)={\rm tanh} (t-t_0)$  
solution, where if we fit a boundary condition $f(0)=1-\epsilon$, we find $t_0=-\log \epsilon$.

Looking at (\ref{singleeq}) the reader might say,
``Of course something
drastic will happen as we pass through the value $\gamma=1$ from below to above. The
potential in the equation now goes to $- \infty$  as $x \rightarrow \pm \infty$. The underlying
field theory must lack a ground state." But we should emphasize that in the context in which
(\ref{singleeq}) was derived, first, the input physics always limits values to numbers determined
by the neutrino density. 
\section{Four groups of neutrinos with standard couplings}
The last exercise was interesting, although not directly applicable to neutrinos themselves, since
for the neutrino case we have $\gamma=1$, with an SU2 invariance in the flavor space. 
But since this case lies on the boundary
between the stable \footnote{i. e., stable in our defined ``short term" sense over the time period of order $T_S$.} and
the unstable case, it is not surprising that with a more complicated initial configuration we
find parameter regions in which the instability is present when we use the usual neutral
current couplings. To illustrate we consider the equation for the densities, $\vec P_A,\vec P_B, \vec P_C,\vec P_D$, 
for the case of four
groups. In the two group case, the angular dependence of the $A-B$ interaction was subsumed in
an effective coupling constant, $g$. In the four group case we define six effective couplings 
$g_{A,B}= \sqrt 2 (1-x_{A,B})n_\nu G_F$, etc.  The equations 
of evolution are then,
\begin{eqnarray}
{d \over dt} \vec P_A =-g_{A,B} \vec P_A \times \vec P_B-g_{A,C} \vec P_A \times \vec P_C
\nonumber\\
-g_{A,D} \vec P_A \times \vec P_D\,,
\label{4group}
\end{eqnarray}
plus six more equations for $(d/dt) \vec P_B$ and $(d/dt) \vec P_C$, obtained by permuting subscripts. The
functions $\vec P_D$ can be eliminated through,
\begin{equation}
\vec P_D +\vec P_A+\vec P_B+\vec P_C={\rm const.}~,
\end{equation}
where the constant depends on the initial conditions.

In fig. 3 we show numerical plots for the time evolution of  $P_A^{(3)} $ when the four
groups consist of two that are respectively in the $\pm z$ directions, and two others that are in
opposed directions at a very small angle to the $\pm z$ directions. 
\begin {figure}[ht]
    \begin{center}
        \epsfxsize 2.75in
        \begin{tabular}{rc}
            \vbox{\hbox{
$\displaystyle{ \, { } }$
               \hskip -0.1in \null} 
} &
            \epsfbox{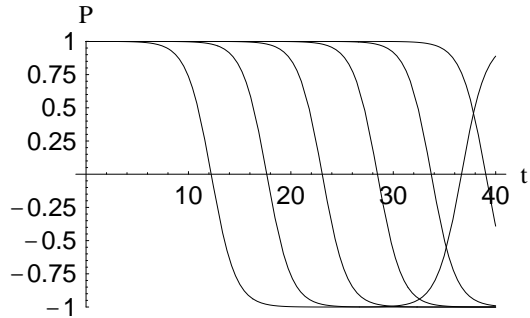} \\
            &
           \hbox{} \\
        \end{tabular}
    \end{center}
 \label{fig.3}
 \protect\caption
    {%
The flavor turnover of group A, from the solution of eqns. (\ref{4group}), where we have taken
the cosines of the angles between the groups as, $x_{A,B}=-1, x_{A,C}=.9, x_{A,D}=-.9,
x_{B,C}=-.9,x_{B,D}=.9,x_{C,D}=-1$. Initial flavors were taken as $r$ for groups A and C, and as
$s$ for groups B and D. Then a small rotation of angle $\epsilon$ around the $(2)$ axis in flavor space was
performed for the initial conditions of group B only. Shown are the curves of $P\equiv P^{(3)}_A$
for values $\epsilon=.01,.001,.0001,.00001,.000001,.0000001$.
 }
\end {figure}
The initial assignment of flavor
for the four groups consists of two sets of flavor $r$ and two of flavor $s$, 
are explained in more detail in the captions. As in the examples in the previous section, we took some flavor mixing, of order  
$\epsilon$, in the initial state, and plotted curves for equally spaced values of $(-\log \epsilon)$. 

We see a result very much like that of the previous section for the case of $\gamma=0$. The time required to reach the first
turn-over, in which $\langle P_A^{(3)} \rangle $ reaches the negative of its original value appears again to be proportional to
$(-\log \epsilon )$. Now, however, we are dealing with the correct coupling scheme, with its
full rotational symmetry in the internal space.
We have explored the space of input parameters (in numerical 
calculations) sufficiently to say that there exists a substantial region of parameter space in which
the behavior is generically like that shown in figs. 3, as well as a region of parameter 
space in which there is only very tiny change in occupancies over the time interval shown in
the figures. In the latter cases the changes remain quadratic in
$\epsilon$ for small $\epsilon$ . We do not have an
analytic reduction of the equations that provides us with the function of the parameters that defines
the boundary of the region, as we had in the broken-symmetry two group case parameterized
with $\gamma$. From the perspective of the long time-scale, that is to say, the usual neutrino 
oscillation time-scale, we would describe the cases where anything significant happens over the time $t<<T_L$ as
being unstable.

Looking more closely at the possible physics in the situation that we have simulated, the reader who
puzzles through the angular domains described in the caption to fig. 3, will see that we began
with two groups moving in nearly the same direction (say the $\hat z$ direction),    one with flavor $r$, the other with
flavor $s$, and two more groups moving nearly in the opposite direction, also with
one flavor $r$ and the other flavor $s$. Then the reader might ask, ``If the alleged turnover
just changes all $r$'s to $s$'s and vice versa, then it hasn't done much, has it? We still have
a group of $r$'s and a group of $s$'s moving more or less in the $\hat z$ direction, and 
a group of $r$'s and a group of $s$'s moving roughly in the $-\hat z$ direction as well."  However, if the \underline{energy
spectra} of the original $r$ groups was greatly different from that of the original $s$ groups,
then we have indeed made a major change, since after the turnover, the $s$ flavors have the
energy spectrum of the original $r$ flavors, and vice versa. We also note that by choosing other geometries
and parameters we can contrive situations in which there is gross change in the angular
distribution for a a particular flavor (though, of course, absolutely no change in the total
angular distribution for neutrinos).

Pending resolution of
questions about the behavior of systems with more continuous distributions in angle,
which we discuss a little more in the next section, 
we shall not address the complexities of application of these ideas to 
the environment of the supernova, where there is much anisotropy. However we should point out that the
possible interesting unstable behavior can be expected to occur on a faster time scale,
and farther into the supernova core, than in a theory with only istropic distributions.We note that the
calculations presented in ref. \cite{sig}, where a flux-weighted angular average of $(1-\cos \theta_{p,q})$
is used, do not capture the physics of the present paper.
Though it takes note of the angular factor, this approach reduces the evolution equation
exactly to the form of the isotropic case with an effective coupling constant, and therefore removes the possibility 
of finding the rapidly growing modes that are the focus of the present paper.
Moreover, it is in the vicinity of the neutrino-sphere that the angular distributions
of the electron neutrinos and of the mu or tau neutrinos are very different, giving, it would
appear, the maximum possibility for rapid spectral and angular exchanges to occur.

A further caveat for the supernova case is that we have absolutely no insight into the
nature of our instabilities in a world in which three flavors of neutrinos are present. This
question calls for further examination.

\section{5. Linear stability analysis when the initial states are flavor eigenstates.}
We have presented two examples in which a small flavor-mixing perturbation
in the initial condition leads to exponential growth followed by nonlinear oscillations at early
times. In these examples, the neutrino distribution in momentum space was rather artificially limited 
to a small number of discrete directions in this space. 
Ideally we would now examine the behaviors
that are possible when the initial flavor distribution in momentum space is more continuous.
Direct numerical simulations over the full range of the non-linear oscillations are ruled out by their complexity. But we
note that in the discrete-angle examples of the last sections the early exponential growth
period could have been studied in the approximation of keeping only the linear response
to perturbations of the initial conditions.
We have not even been able to carry this programme through in any general way for the case of continuous distributions in angle, 
since the non-linearities defeat the partial
wave expansion that is a natural beginning point. However we can use a partial wave expansion
for the case in which the unperturbed solution is isotropic, in order to confirm our belief
that there are in this case no growing modes of any multipolarity.

We begin with a set of neutrino density functions $\vec Q(\Omega)$ defined for each direction
of space (i.e. the momentum space of the neutrinos), 

\begin{equation}
\vec Q(\Omega) \, d \Omega ={1 \over n_\nu V} \sum_{i \subset \{ d \Omega\} }\vec \sigma _i ~,
\end{equation}
where the notation indicates that the sum is to be taken over all states for which the direction of momentum
lies within the element of solid angle $d \Omega$.
We introduce a second set of densities, $R(\Omega)$, for some other neutrino states in the ensemble, for which
 the one-particle operators are labeled as $\vec \sigma_{i'}$,
\begin{equation}
\vec R(\Omega ) \, d \Omega ={1 \over n_\nu V} \sum_{i' \subset \{ d \Omega\} }\vec \sigma_{i'}~ .
\end{equation}

The
physical distinction between $\vec Q(\Omega)$ and $\vec R(\Omega)$ could be that they represent different, disjoint
parts of the energy spectrum (which could have differing initial flavor occupancies in some applications.) The notation
$i'$ would then indicate that the sum was over a set of energy states disjoint from those indexed with $i$. Alternatively, 
the $R(\Omega)$ variables could represent the densities of antineutrinos in which case $i'$ designates the 
operators for antiparticles. The way in which anti-particle densities enter the equations 
which we shall write has been treated in detail in ref. \cite{prs}, and in several other references already cited,
although with some variations in the definitions of the density operators. We use the negative of the density matrix elements
defined for antineutrinos in ref \cite{sr}, making the resulting structure completely symmetrical between neutrino
and antineutrino (see Appendix A). The commutation rules, both for the case in which $Q$ and $R$ are distinguished by spectrum
and in the case where they are distinguished by particle-antiparticle classification, are then, 
\begin{eqnarray}
 [Q_i(\Omega)\, ,  Q_j (\Omega ' )]={2 i\over n_\nu V} \sum_k \epsilon_{i,j,k}\delta (\Omega -\Omega ') Q_k(\Omega)~,
\end{eqnarray}
and
\begin{eqnarray}
[R_i(\Omega )\, ,  R_j (\Omega ' )]={2 i \over n_\nu V}\sum_k \epsilon_{i,j,k}\delta (\Omega -\Omega ') R_k(\Omega)~,
\end{eqnarray}
and the equations of motion, in complete analogy to (\ref{eom2}) are now,
\begin{eqnarray}
{d \over dt} \vec Q(\Omega ,t) =g\vec Q(\Omega ,t) \times \Bigr \{ \int d\Omega ' [\vec Q(\Omega ' ,t) 
\nonumber\\
+\vec R(\Omega' ,t)]
(1-\cos \theta _{\Omega,\Omega '}) \Bigr \} ~,
\label{peqn}
\end{eqnarray}
and
\begin{eqnarray}
{d \over dt} \vec R(\Omega ,t) =g\vec R(\Omega ,t) \times \Bigr \{ \int d\Omega ' [\vec Q(\Omega ' ,t) 
\nonumber\\
+\vec R (\Omega' ,t)]
(1-\cos \theta _{\Omega,\Omega '}) \Bigr  \} ~ .
\label{qeqn}
\end{eqnarray}

We first consider the application of (\ref{peqn}) and (\ref{qeqn}) to a system that is totally isotropic
(in momentum space) in its unperturbed state, as well as being initially in a flavor eigenstate. 
We write the two densities as
\begin{eqnarray}
\vec Q(\Omega,t)=\vec Q_{(0)} +\sum_{l,m} Y_{l,m} (\theta, \phi) \vec q_{l,m}(t) ~,
\nonumber\\
\vec R(\Omega,t)=\vec R_{(0)} +\sum_{l,m} Y_{l,m}  (\theta, \phi) \vec r_{l,m}(t) ~,
\label{pertdef}
\end{eqnarray}
where we take space-isotropic and flavor-diagonal values for the unperturbed densities,
\begin{eqnarray} 
\vec Q_{(0)}=a \hat k   \,\,;\,\, \vec R_{(0)}=b \hat k 
\label{zero}
\end{eqnarray}
with $\hat k$ a unit vector in the (3) direction in the internal space,
The perturbations have been expanded in spherical harmonics (in the momentum space) with 
coefficients $\vec q_{l,m}, \vec r_{l,m}$.
With these choices, the equations (\ref{peqn}) and (\ref{qeqn}) are obeyed if we set $\vec q_l(t)=0$ and $\vec r_l(t)=0$.
Thus, the linearized equations for the time dependence of the perturbations are homogeneous, with
time independent coefficients. They follow directly from substitution of (\ref{pertdef}) into
(\ref{peqn}) and (\ref{qeqn}), yielding, for the cases $l=0$, and $l=1$,
\begin{eqnarray}
g^{-1}{d \over dt}\vec  q_{0,0}(t)=-{d \over dt} \vec  r_{0,0}(t)
\nonumber\\
=[- a \vec r_{0,0}(t)+b\vec q_{0,0}(t)] \times \hat k ~,
\end{eqnarray}

\begin{eqnarray}
g^{-1}{d \over dt}\vec  q_{1,m}(t)=[({4 a\over 3}+b)\vec q_{1,m}(t)
+{a} \vec r_{1,m}(t)] \times \hat k,
\label{stab1}
\end{eqnarray}
\begin{eqnarray}
g^{-1} {d \over dt}\vec  r_{1,m}(t)=[({4 b\over 3}+a)\vec r_{1,m}(t) 
+{b}\vec q_{1,m}(t)] \times \hat k ~ .
\label{stab2}
\end{eqnarray}
For values $l > 1$ we obtain.
\begin{eqnarray}
g^{-1}{d \over dt}\vec  q_{l,m}(t)=
(a+b) \vec q_{l,m}(t) \times \hat k
\nonumber\\
g^{-1}{d \over dt}\vec  r_{l,m}(t)= 
(a +b) \vec r_{l,m}(t) \times \hat k ~.
\label{stab3}
\end{eqnarray}
It is easy to see that in this case, with isotropic unperturbed distributions, all of the eigen-frequencies of  (\ref{stab1})-
(\ref{stab3}) are real. Note that the complication due to the multiplicity the vector components, $\vec q$, $\vec r$ in the above
equations is superficial. The eigen-frequencies of the system (\ref{stab1}), (\ref{stab2}) are given exactly by the eigenvalues of the
matrix, 
\begin{equation}
g \pmatrix{{4\over 3} a +b &a\cr
	b&{4 \over 3}b +a\cr} ~.
\end{equation}
If we had introduced angle dependence into the zero'th order distributions (\ref{zero}),
the linearized equations for the perturbations would still have time independent coefficients, but all
values of $l$ would have been coupled together. We therefore arrive at no conclusion as to the presence of
growing modes in this case.

\section{6. Quantum case}

We review the development to the present point. The original Hamiltonian for several groups
of neutrinos is a sum of pairwise interactions among collective operators for each group. In the Heisenberg
equation of motion for these operators we then replaced the operators by  their expectation values, 
and we replaced products of the operators by products of expectation values, following assumption (a) of sec.1.
We considered only initial states for which this assumption is true, but the assumption cannot remain exactly true
as time progresses. The applications of the classical equations proposed in the literature demand that this classical approximation
be good over the longer timescale $T_L$. A heuristic justification might be that when the number of particles in each
group is large, then the collective operators become, in effect, the operators for a system with very large spin. We are accustomed to
the phenomenon of large quantum numbers leading to classical behavior. One of our aims in the present paper
was to test the classical approach by comparing with the direct numerical solution of the models, the
latter requiring up to several hundred spins, in order to estimate the dependence on the particle number.
 Here we address this question through numerical solution of the complete Schrodinger equation, for the
the  simpler configurations considered above. In appendix B, using the results of this section, we shall
further illuminate the use of the term ``quantum" in distinguishing this approach from that of the previous
sections.

We begin with the simplest interesting example, the two group model (\ref{ham4}), where the internal SU2
symmetry is broken by taking $\gamma=0$, but where we now start with an exact flavor eigenstate, with the parameter of the previous calculation, $\epsilon $, taken as zero. In the initial state the
neutrinos in the upper group all carry the $r$ flavor and the neutrinos in the lower group all carry the $s$ flavor.
Results are shown in fig.4. 

\begin {figure}[ht]
    \begin{center}
        \epsfxsize 2.75in
        \begin{tabular}{rc}
            \vbox{\hbox{
 $\displaystyle{ \, { } }$
               \hskip -0.1in \null} 
} &
            \epsfbox{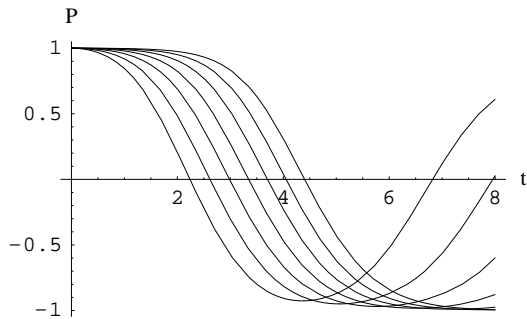} \\
            &
            \hbox{} \\
        \end{tabular}
    \end{center}
 \label{fig.4}
\protect\caption
    {%
Curves for the model of sec. 3, but now based on a complete quantum mechanical solution
using the Hamiltonian, and with no ``seed" of flavor mixing in the initial condition. Now the
turn-over effect comes about quantum mechanically and is dependent on the particle number
$N$. We plot $P$ as a function of scaled time, for values $N=8,16,32,64,128,256,512$.
The curves are to be compared to those of fig. 2.
 }
\end {figure}

Comparing with the plot of fig.2, which showed the time development for the classical case, but with $\epsilon \ne 0$ we find a surprising similarity. We can state the result of the comparison as follows:
 in the classical case, seeded with an initial perturbation that is of order $\epsilon$, the time for
flavor turn-over is of order $(-T_S \log \epsilon)$; in the quantum case \footnote{As we saw, $N$ is not an explicit parameter in the equations for the $\vec P$'s in the classical case; it was removed from the equations of motion when the intensive density operators $\vec P_a$ and the coupling parameter $g$ were introduced.}, with no seeding of initial flavor
mixing, the turn-over time is of order $T_S \log (N)$, where $N=n_\nu V$ is the number of neutrinos. The shape
of the turn-over curves is almost identical.

Carrying out the same calculation for the case $\gamma \ne 0$ leads to a family of curves much like those
shown in fig. 2, again with the $\log N \rightarrow (-\log \epsilon )$ correspondence.
In the solutions of the complete problem based on  (\ref{ham4}) the dependence on $N$ is twofold;
first the implicit dependence through the coupling parameter $g=2 \sqrt 2 G n_\nu$ and second, the residual
explicit dependence that enters the above result. It is this residual dependence that shows the logartihmic behavior
in $N$ for large values of $N$.

We have also looked at the direct solution of the evolution under influence of the complete forward 
Hamiltonian for the example of four neutrino groups that was treated classically in sec. 4.  The results are
completely in accord with our expectations derived from the above, namely, turn-over times that
are of order $T_S \log N$, and the relation $(-\log \epsilon ) \leftrightarrow \log N$.
We cannot present quite as good curves of the $N$ dependence in this case because of computational
constraints. The method of direct solution takes advantage of the fact that each group of neutrinos
is represented by total spin operators for that group; thus with a given total number of neutrinos
the dimensionality of the calculation goes up with the number of groups. Still the calculation lends
support to the conclusion that the relation of classical instability to quantum speed-up is
generic.

In applications to supernova neutrinos we would usually expect that $ (-\log \epsilon) >>\log N_{\rm eff}$, where we estimate 
$\epsilon \approx T_S \delta m^2/E$, and $N_{eff}$ is the number of neutrinos in a volume of dimension $c T_S$. Thus 
the classical approach based on (\ref{eom4}) would appear to be sufficient in the 
supernova application. The spontaneous effect, i.e. the transformations
that the neutral current interactions can bring about in the absence of a flavor-mixing term, in the absence of flavor
mixing in the initial state,
will be small compared to the seeding effects of the mass$^2$ terms, which produce the small mixing of order $\epsilon$
that in turn triggers growth of the classically unstable mode.
It would be worth confirming this conclusion in detailed simulations, but that is beyond the scope of
the present paper. 

We mention that there exists a system which, in effect, is governed by a model that is isomorphic to the model of
sec. 3 with parameter $\gamma=0$, namely the case of the encounter of two photon clouds,
where one or both of the clouds has large circular polarization, and where the interaction is the given by the standard
Heisenberg-Euler formula. In ref. \cite{rfs2} it is shown that large polarization exchange can occur over
times that are many orders of magnitude smaller than one would estimate from ordinary photon-photon
cross-sections, in analogy to the rapid flavor exchange that we find in the above models.

We are now in a position to state a preliminary conclusion with respect to assumption (a) discussed
in sec. 1. We indeed have evidence that when the number of particles approaches $\infty$, the 
expectation value factorization, generically $\langle \vec \sigma_i \times \vec \sigma_j \rangle=\langle \vec \sigma_i \rangle \times \langle \vec \sigma_j \rangle$, if present initially, remains true for all times on the short time scale.
However,  this statement holds only up to the $(\log N )^{-1}$ corrections to reaction rates.

These relations between 
the numerical calculations from the complete Hamiltonian and the results in the ``classical" approximation
can be elucidated in our simplest two-group case in an analytic approach.
Again choosing the model of sec. 3, with the Hamiltonian (\ref{ham4}) and taking $\gamma=0$, we introduce a set of operators that are bilinear in the variables $\tau$ and $\zeta$,

\begin{eqnarray}
x=i\tau^{(+)} \zeta ^{(-)}~;~ u=i\tau^{(-)} \zeta^{(+)} ~;~  y=\zeta^{(+)} \zeta^{(-)},
\nonumber\\ z=\tau^{(-)} \tau^{(+)} ~~~~;~~~~~ w=\tau^{(3)} ~,
\label{fiveops}
\end{eqnarray}
where $\tau^{(+)}=\tau^{(1)}+i \tau^{(2)}$ etc.

We now write Heisenberg equations of motion for these operators by taking commutators
with $H_{\rm for}$ of (\ref{ham4}), specializing to the case, $x_{A,B}=0$, and taking the conserved quantity $\tau ^{(3)}+\zeta^{(3)}=0$,  We measure
time in units $T_S=(G_F n_\nu)^{-1}$
and obtain the closed set,
\begin{eqnarray}
&N\dot x=w(z+y)+w^2 ~~;~~\dot u=-\dot x ~~;~~N \dot y=w x-u w,
\nonumber\\
&N \dot z =xw-wu~~;~N \dot w=-x+u ~.
\label{quad}
\end{eqnarray}
Now we take (\ref{quad}) as c-number equations for the expectation values of $x,y,z,u,w$ .
We choose the initial system to have $N$ states in group $A$ all occupied with neutrinos of flavor $r$, 
and $N$ states in group B, all occupied by neutrinos of flavor $s$. Then the initial values of the variables defined
above are $x=y=z=u=0$ and $w=N$.
For this case the set (\ref{quad}) leads to a single second-order equation for $\bar w \equiv w /N$, 
\begin{equation} 
{d^2 \over dt^2} \bar w =2 \bar w( \bar w^2 -1) -{2 \bar w^2 \over N} ~.
\label{soliton}
\end{equation}
where the initial condition is now $\bar w=1$. In appendix B we give a fuller explanation of the above steps, and in addition, we
restore $\hbar$ to the calculation to show the sense in which the last term on the RHS in (\ref{soliton}) is a quantum
correction to the classical approximation.

If, instead of using the equations (\ref{quad}) for the bilinear forms, 
we had used the Heisenberg equations for $\vec \tau$ and for $\vec \zeta$,
taken now as classical quantities, and only then had introduced the variables $x,y,z,u,w$ as defined above,
we would have retrieved the set (\ref{quad}) with one modification; the first equation would now be just,
$N\dot x=w(z+y)$, and (\ref{soliton}) would be replaced by ${d^2 \over ds^2} \bar w =2 \bar w( \bar w^2 -1) $. This set now lacks the term that drives the flavor turnover in time $T_S \log (N)$. Indeed, as remarked earlier, it has no $N$
dependence. 

We can derive the logarithmic behavior analytically from (\ref{soliton}),
capitalizing on the fact that when $N\rightarrow \infty $, the solution is the familiar $\bar w={\rm tanh} [(t-t_0)]$ 
kink solution in a $\lambda \phi ^4$ theory in one dimension, then showing that for large $N$,
in the time region in question, the $ \bar w^2 / N$ term in (\ref{soliton}) can be dropped in favor
of changing the initial value of $\bar w$ to $1-2/N$,  this in turn determining $t_0=\log(2N)$. 

\section{7. Discussion}

We began with the standard classical equations of motion for neutrino flavor densities under conditions in which
the time-scale of the neutral current interaction $T_S$
is short compared to the flavor oscillation time-scale $T_L$.  Our first object was to examine solutions of these
classical equations for the case
of initial conditions that are flavor-diagonal, or nearly so, but with flavor-dependent spatial anisotropies of the
initial momentum distributions of the neutrinos. We then solved for the time scales for significant change in the distributions. 
Although our particular solutions are periodic, in any physical context with randomized initial conditions 
these time scales would be indicative of the time required for flavor-equilibration. 

In our actual calculations based on the classical equations we discarded the neutrino mass term 
(which generates conventional oscillations) in favor of initial conditions that depend on a parameter $\epsilon$ which
is to be interpreted as the amount of neutrino mixing that develops, starting from a flavor-diagonal
state, on the short time-scale $T_S$.

Depending on the details of the initial distributions,
 we found two dramatically different possible behaviors,
separated by paper-thin transition zones. In one case there is essentially no evolution of
occupancies over the short time scale $T_S$. In the second case there is large evolution on the time-scale 
$(-T_S \log \epsilon)$. The simplest model that displays this transition is the two-group
model of sec. 3 where we changed the neutral current coupling somewhat by introducing
the parameter $\gamma$, and where the physical value $\gamma=1$ turns out to be exactly the boundary of short-term stability. But in the multigroup cases that we studied, there
are regions of instability for the physical case $\gamma=1$, as well.

Our second objective in this paper was to better understand the limitations of the classical approximation 
that underlies almost all of the literature on this subject. We recall that our path
to these equations involved two steps: 1) Replacement of the complete neutral-current Hamiltonian
by a forward Hamiltonian that includes only the momentum modes occupied (by neutrinos
of either flavor) in the initial state ; 2) The factorization assumption (\ref{uncor}). In sec. 6
we investigated this latter assumption by direct computations of evolution beginning from
the forward Hamiltonian. In order to achieve the maximum clarity in the interpretation of results, 
we began from exact flavor eigenstates, $\epsilon=0$, in carrying out these solutions. For this case,
of course, the classical equations give no evolution.  Under conditions for which the
classical equations are stable, in the sense defined above, the numerical solutions for the quantum case, indeed show only
a tiny evolution over the time scale $T_S$, by an amount that decreases (as $N^{-1}$) as $N$ is increased.

However when we look at the same comparison under conditions in which the classical solutions
are unstable, the results are very different. In this case the numerical solutions for finite $N$ can undergo complete
turnover of flavors, from one part of momentum space to another, in a time that is apparently of order $T_S \log N$. 
The interesting
comparison is now to the classical approximation, now with a small value of $\epsilon$, where
we found similar development over the time scale $(-T_S \log \epsilon)$.
These observations are based on simulations that are, for all cases but one, a bit inadequate in
covering the range in $N$ required to make a firm conclusion as to the logarithmic behavior; hence our
use of the word, ``apparently", in the above.

In the simplest model that we considered, that of section 3, we were able both to calculate
for large enough values of $N$ to make the above conclusion fairly firm, and also to find an analytic
approach that supports the logarithmic behavior. Qualitatively we could describe this latter approach as 
being the next step in a heirarchy of approximations that begins with the classical approximation.
The classical approach writes equations for densities, quantities bilinear in the neutrino fields, with a factorization assumption to close the set. This is now replaced with equations for quantities that are bilinear in the densities, again with a factorization assumption to close the set. The solutions of this second set of equations well
replicate the results of the numerical study of the quantum case over our whole range of parameters.

From these results, and further simulations that we have not presented here, we have what we believe is
a generic relationship between the instabilities of the classical approximation and the 
$T_S \log N$ behavior of the full models. Then returning to item \# 4 on the list of possible neutrino
collective phenomena at the beginning of this paper, we have at least a partial answer to
the question of whether there is ever a ``speed-up" of the sort proposed in ref. \cite{bell} but questioned
in ref. \cite{fl}. Ref. \cite{bell} addressed the case with an initial condition in which 
the two neutrino flavors occupied different spectral regions but in which the angles between
the momenta were distributed over the whole region $0- 2\pi$, and asked for the characteristic time
for the spectral rearrangement. The conjecture presented was that the evolution would be at the
``speeded-up" rate characterized by $T_S$. Because of the scatter in couplings, as well as inadequate
numerical methods, ref. \cite{bell} achieved simulations with a maximum value of only N=14, far too small
a value to find the logarithm if that were the answer. In the present paper, we have been able to study cases with
much higher values of $N$ through taking a few beams narrow in solid angle, and we find the speed-up discussed 
above, for certain initial conditions.
\footnote{Note that even with the $\log N$ factor included we can legitimately claim a ``speed-up" in many
circumstances, since $T_L$ is often many orders of magnitude greater than $T_S$.} Ref. \cite{fl}
took all couplings to be the same, as for the totally isotropic ``classical" equation, and solved the
full problem analytically, finding no ``speed-up". Since the classical isotropic case is stable,
as we saw in detail in sec. 5, the lack of a speed-up in the full solution to the model of ref. \cite{fl} model presents no inconsistency 
with our generic relation between stability of the classical case and speeded-up behavior
for the full model.

There is a difficult question that we have not addressed here. We have studied systems that consist of plane
waves that fill our quantization volume. In our models the participating particles lie on top of each other for the whole time
interval over which we study the system. In actual physical systems that we might consider, the particles moving
in a given direction must have some characteristic length that determines how long they stay in contact
with those moving in the opposite direction. We are presently not in a position to say exactly how
this affects the analysis of the ``quantum" evolution effects in any realistic situation (we believe, as do
other authors, that it affects the ``classical" evolution not at all). We can try to gain insight with calculations
which are similar to those reported in sec. 6, but where instead of having two groups 
of particles in contact with each other over our complete time interval we focus on the state of 
a single set, moving in the same direction, and take this set to be serially in contact with a number of groups
of other particles, one group at a time. We can state some prelimary results of these calculations in a negative form,
first by posing the question, ``As our test group is affected first by one other group, and then by another, and by another,
is it just a question of determining, after each encounter, the probabilities of transitions in the test
packet, and then proceding to the next encounter with an flavor-diagonal ensemble embodying
these probabilities?" If the answer were yes, then of course there would be no evolution beyond
that which is be achieved in a single short encounter time, multiplied by the number of interactions.  
Based on simulations that we have done, reinforced 
with some analytical considerations, we can state firmly that the answer to the above question is ``no". 
There is a coherent
effect, in spite of the independence of the colliding packets, which gives rate enhancements. We shall
return to this subject in a future publication.

There are also possible questions relating to our use of the ``forward" Hamiltonian (\ref{ham})
rather than the full Hamiltonian for the system; that is, to the neglect of all but forward scattering
processes. Here we believe that we are on a firm footing, as long as we
consider times that are very much less that $T_{\rm scat}$. Inclusion of the non-forward parts gives
a forest of oscillating terms that adds up essentially to nothing over this time scale.
We know of no reference that treats such questions in detail, but we do note that the whole literature
on what we have called the classical approach is built on th1s assumption.

\section{Appendix A. Inclusion of antiparticles.}
Since the above work did not explicitly deal with neutrino-anti-neutrino mixtures, and such
mixtures will enter any application to which the formalism is relevant, we show how anti-particles fit into the general framework without changing the structure of the equations.
We define the operators,
\begin{eqnarray}
  \sigma^{(+)}_j=(a^r_j)^\dagger a^s_j +(\bar a^s_j)^\dagger \bar a^r_j ,
\nonumber\\
\sigma^{(-)}_j =(a^{s}_j)^\dagger a^{r}_j+(\bar a^{r}_j)^\dagger \bar a^{s}_j ,
\end{eqnarray}
where now the barred operators create and annihilate the anti-neutrino states for flavor $(r,s)$ and momentum $p_j$.
We note the relation,
\begin{eqnarray}
[ \sigma ^{(+)} _ j , \sigma ^{(-)} _ j ] = \sigma^{(3)} _ j =
(a^{r}_j)^\dagger a^{r}_j - (a^{s}_j)^\dagger a^{s}_j \, 
\nonumber\\
-(\bar a^{r}_j)^\dagger \bar a^{r}_j +(\bar a^{s}_j)^\dagger \bar
a^{s}_j ~.
\label{anti}
\end{eqnarray}
Now forming the combinations, $\sigma^{(1)}_j=(\sigma^{(+)}_j+\sigma^{(-)}_j)$, $\sigma^{(2)}_j=-i(\sigma^{(+)}_j-\sigma^{(-)}_j)$
for each mode, and substituting in (\ref{ham}), where the index $j$ labels momentum only, we obtain the complete \underline{relevant} part of the (neutral-current
plus oscillation) forward Hamiltonian, including antiparticle operators, as can be shown by a tedious calculation.
By relevant we mean that all the remaining interaction terms,
as well as the kinetic energy terms, commute with each of the $\vec \sigma_i$. The $\vec \sigma_i$ operators, as extended
above, still obey angular momentum commutation rules, so that all of our previous formalism for following the distributions in time
is applicable to groups that contain mixtures of particles and antiparticles, with the provision, of course, that 
the total amount of one particular flavor is counted by subtracting the number of anti-particles of that flavor from the
number of particles, as in (\ref{anti}). Thus the equations (\ref{eom}) still describe the time evolution of the
system.

We may wish to separate particles from antiparticles, however, putting them in disjoint groups 
with collective operators $\vec P_a$, where various values of the subscript $a$ now both distinguish
energy bins and distinguish particles from antiparticles, and where an anti-particle operator is defined, for example, as,
\begin{equation}
 P_a^{(3)}={1\over n_\nu V}\sum_{j \subset \{ a\} }\langle  -(\bar a^{r}_j)^\dagger \bar a^{r}_j +(\bar a^{s}_j)^\dagger \bar
a^{s}_j \rangle ~.
\end{equation}
Then it is easy to see that dividing the collective 
operators into particle and antiparticle parts indexed with different $a$
preserves the commutation rules of the underlying operators, and that the equation of motions for the $P_a$'s, after we make
the classical factorization assumption  (\ref{uncor}), are again given by (\ref{eom2}),

\begin{equation} 
{d\over dt} {\vec P_a}=\lambda_a {\vec B}\times {\vec P_a}- g   \vec P_a
\times  \sum_{a,b} (1-\cos \theta _{a,b}) \vec P_b ~.
\label{eomf}
\end{equation}

We contrast this equation with the equations that have been given in \cite{prs} and other sources, here specialized to
the isotropic case where (in effect) we have $\cos \theta _{a,b}=0$,
\begin{eqnarray}
{d \over dt} {\bf  P}_j={\Delta m^2 \over 2 p_j} {\bf B \times P}_j + {\sqrt{2} G_F \over V}{\bf P_j}\times \sum_i ({\bf  \bar P}_i-
{\bf  P}_i)\, ,
\nonumber\\
{d \over dt}{\bf \bar P}_j=-{\Delta m^2 \over 2 p_j} {\bf B \times \bar P}_j + {\sqrt{2} G_F \over V}{\bf \bar P_j}\times \sum_i ({\bf  \bar P}_i-{\bf  P}_i)
\nonumber\\
~
\label{prs}
\end{eqnarray}
where $\bf P_j$ stands for a particle
density (or density matrix element) and $\bf \bar P_j$ stands for an antiparticle density.

We can reconcile (\ref{eomf}) with the set (\ref{prs}) first by taking into account the difference in normalization 
of the $P_a$'s in the two cases, and the corresponding relation of coupling constants. \footnote{Our normalization
is such that $\sum P^{(3)}_a=(n_r-n_s)/ n_\nu$; that of ref. \cite{prs}has the additional factor $n_\nu$
on the RHS.} and then following with
the replacement ${\bf \bar P}_i \rightarrow -{\bf \bar P}_i$ in the equations (\ref{prs}), from ref \cite{prs}. The latter 
change reflects a difference in definitions of the densities\footnote{As noted in ref. \cite{prs}, much of the 
literature uses definitions that give equations that are even less symmetrical between particle and antiparticle than (\ref{prs}). 
We refer the reader to ref.\cite{prs} for details.}
Our definitions are better suited to our purposes because when we consider the operators of which the
$P$'s are expectation values, it is with our choice that both the particle and the antiparticle sets of underlying operators
have the same commutation rules (angular momentum commutation rules modified by the factor $(n_\nu V)^{-1}$).

\section{Appendix B. Illustration of the comparison of quantum and classical calculations.}
When we introduce $\hbar$ into the problem and define the commutation rules for the Fourier components
of the fields in the canonical way, then forming the bilinear forms $\vec \tau$ and $\vec \zeta$ as in (\ref{operators}),
the commutators become $[\tau^{(1)}_i,\tau^{(2)}_i]=2 i \hbar \tau^{(3)}_i$ ,etc. Taking $X_{A,B}=0$, $\gamma=0$, the Hamiltonian (\ref{ham4}) is  

 \begin{equation}
H_{\rm for}={G_F  \over \sqrt 2 V \hbar ^2}( \tau^{(1)}  \zeta ^{(1)}+\tau^{(2)}  \zeta ^{(2)}) .
\label{scham2}
\end{equation}
As in sec. 6 we define
\begin{eqnarray}
x=i\tau^{(+)} \zeta ^{(-)}~;~ u=i\tau^{(-)} \zeta^{(+)} ~;~  y=\zeta^{(+)} \zeta^{(-)},
\nonumber\\ z=\tau^{(-)} \tau^{(+)} ~~;~ ~w=\tau^{(3)} ~,
\end{eqnarray}
then calculating $i \hbar^{-1} [x,H]$, and obtaining (using unscaled time),
\begin{eqnarray}
\dot x= {G_F \over V \hbar^2}[\tau^{(3)} \zeta ^{(+)} \zeta^{(-)}- \zeta^{(3)} \tau^{(+)} \tau^{(-)}]
\nonumber\\
= {G_F \over V \hbar^2}[\tau^{(3)} (\zeta ^{(+)} \zeta^{(-)}+ \tau^{(-)} \tau^{(+)})+\hbar (\tau^{(3)})^2]
\nonumber\\
={G_F \over V \hbar^2} [2 w z + \hbar w^2] ~.
\label{xeqn}
\end{eqnarray}
To get the second line we have here used $\tau^{(3)}+\zeta^{(3)}=0$ and evaluated one more commutator in order to get the 
product of $\tau$'s in the order that agrees with the definition of $z$, above. Note that some guile has been
used to get a closed set of equations for our five quantities, and that the procedure is somewhat arbitrary as well,
since we have used the noncommutative properties of our operators up to the point at which we can reexpress
results in terms of our original variables. Then we replace them by c-numbers that we take to be their
expectation in the medium.  (Note also that the ``classical" approximation that we discussed earlier is subject
to exactly the same criticism.) With our initial conditions $x=y=u=z=0$ and using the equations $\dot x=-\dot u$,
$\dot y=\dot z$ (where in the latter the expectation value ansatz avoids an order-of-operator difference), we obtain
$u(t)=-x(t)$ and $y(t)=z(t)$. Then the two remaining equations of motion are easily found to
be,
\begin{eqnarray}
\dot w ={G_F \over V \hbar^2}[-2 x]~,
\nonumber\\
\dot z={G_F \over V \hbar^2}[2 x w] ~ .
\label{other2}
\end{eqnarray}
We note that the last term on the RHS of the equation for $\dot x$ has the only factor of $\hbar$ in the three
equations of motion, aside from the factor of $\hbar^{-2}$ that multiplies $G_F$, which will enter into any
time-scale depending on $G_F$. Thus, in a well defined sense, this term is a quantum correction. Indeed, if
we had calculated the equations of motion for these bilinear forms using the original classical equations for the
linear forms, (\ref{eom4}) for the present case, the result would have been exactly
the above set, except lacking the last term in the $\dot x$ equation, with its explicit $\hbar$.  Therefore we 
have a basis for calling the classical approximation ``classical" and describing the computer generated solutions
plotted in section 6, or alternatively the solution of (\ref{soliton}), as ``quantum mechanical. "

For the case of our initial conditions, the reduction of the three equations (\ref{xeqn}) and (\ref{other2}) 
to the single equation (\ref{soliton}) for the function $w$ proceeds by noting first that by dividing the first
of eqs. (\ref{other2}) by the second, and doing one integral we obtain,
\begin{equation}
w^2/2=-z+ const.~~.
\end{equation}
Then, reverting to units with $\hbar=1$, scaling the time, choosing $w(0)=N$, and defining $\bar w=w/N$ we 
obtain (\ref{soliton}).


\begin{thebibliography}{00}
\bibitem{wolf} L. Wolfenstein, Phys. Rev. {\bf D17}, 2369 (1978)
\bibitem{ms} S. P. Mikheev and A. Y. Smirnov, Sov. J. Nucl. Phys. {\bf 42}, 913 (1978) 
\bibitem{slow1}R. A. Harris and L. Stodolsky, Phys. Lett. 116B, 464 (1982)

\bibitem{slow2}R. A. Harris and R. Silbey, J. Chem. Phys. 78,7330 (1993)

\bibitem{slow3}P. Silvestrini and L. Stodolsky, Phys. Lett. A280, 17 (2001)
\bibitem{BSV} N.F. Bell, R.F. Sawyer and R.R. Volkas, Phys. Lett. B500, 16 (2001)

\bibitem{s1} S. Samuel, Phys. Rev. D48, 1462 (1993)

\bibitem{s2} V. A. Kostelecky and S. Samuel, Phys. Rev. D49, 1740 (1994)

\bibitem{s3} V. A. Kostelecky, J. Pantaleone and S. Samuel, Phys. Lett. B315, 46 (1993)
\bibitem{s4}V. A. Kostelecky and S. Samuel, Phys. Lett. B318, 127 (1993)
\bibitem{s5}V. A. Kostelecky and S. Samuel, Phys. Rev. D52,3184 (1995)
\bibitem{s6}S. Samuel, Phys. Rev. D53, 5382 (1996)
\bibitem{s7}V. A. Kostelecky and S. Samuel, Phys. Lett. B385, 159 (1996)
\bibitem{s8}J. Pantaleone, Phys. Rev. D58,073002 (1998)
\bibitem{bell} N. F. Bell,  A. A. Rawlinson, and R. F. Sawyer, Phys.Lett. {\bf B573}, 86 (2003) 
\bibitem{rfs}R. F. Sawyer, quant-ph/0312217
\bibitem{fl}A. Friedland and C. Lunardini, Phys.Rev. {\bf D68},  013007 (2003) and
A. Friedland and C. Lunardini,  JHEP {\bf 0310}, 043 (2003) ; hep-ph/0307140. 
\bibitem{sr} G. Sigl and G. Raffelt, Nuc. Phys {\bf B406}, 423 (1993)
\bibitem{prs} S. Pastor, G. Raffelt and D. Semikoz, Phys.Rev. {\bf D65},  053011 (2002)

\bibitem{sn}
S.~Pastor and G.~Raffelt,
Phys.\ Rev.\ Lett.\  {\bf 89}, 191101 (2002).

J.~Pantaleone,
Phys.\ Lett.\ B {\bf 342}, 250 (1995);
Y.~Z.~Qian and G.~M.~Fuller,
Phys.\ Rev.\ D {\bf 51}, 1479 (1995);
\bibitem{earlyu}
A.~D.~Dolgov, et al,
Nucl.\ Phys.\ B {\bf 632} (2002) 363;
Y.~Y.~Wong,
Phys.\ Rev.\ D {\bf 66}, 025015 (2002);
K.~N.~Abazajian, J.~F.~Beacom and N.~F.~Bell,
Phys.\ Rev.\ D {\bf 66}, 013008 (2002).
\bibitem{lunardini}
C.~Lunardini and A.~Y.~Smirnov,
Phys.\ Rev.\ D {\bf 64}, 073006 (2001).


\bibitem{pantaleone}
J.~Pantaleone,
Phys.\ Lett.\ B {\bf 287}, 128 (1992);
J.~Pantaleone,
Phys.\ Rev.\ D {\bf 46}, 510 (1992).


\bibitem{raff}
G.~Raffelt and G.~Sigl,
Astropart.\ Phys.\  {\bf 1}, 165 (1993);
B.~H.~McKellar and M.~J.~Thomson,
Phys.\ Rev.\ D {\bf 49}, 2710 (1994).
\bibitem{sig}G.~Sigl,
Phys.\ Rev.\ D {\bf 51}, 4035 (1995);
\bibitem{rfs2} R. F. Sawyer, hep-ph/0404247


\end{thebibliography}
\end{document}